\documentclass[a4paper]{VisionStyle}
\usepackage{epsfig}

\begin{document}

\title{First XMM-Newton results on the Pulsar Wind Nebula 
in the composite supernova remnant G0.9+0.1}

\author{D. Porquet\inst{1} \and A. Decourchelle\inst{1} 
                           \and R.S. Warwick\inst{2}} 

\institute{Service d'Astrophysique, CEA Saclay, 91191 Gif-sur-Yvette Cedex, 
France
\and Department of Physics and Astronomy, University of Leicester, 
Leicester LE1 7RH, UK }

\maketitle 

\begin{abstract}
 We report the results of the {\sl XMM-Newton} observation of the Pulsar Wind 
Nebula (PWN) in the composite supernova remnant G0.9+0.1 located in the 
Galactic Center region. Thanks to the sensitivity of the EPIC cameras, we 
focus on the first spectral analysis of large and small-scale structures of 
this PWN using MOS and PN data in combination. Our spatially resolved spectral 
analysis at large-scale offers a clear indication of a softening of the photon 
index with distance from the centroid of the nebula 
(from  $\Gamma_{\rm core}$=1.12$^{+0.45}_{-0.48}$ to $\Gamma$=2.42$^{+0.19}_{-0.19}$), 
as observed in other known X-ray plerions. 
A spectral analysis of the small-scale structures in the central region of 
this PWN, delineates variations of the spectral index within the arc-like 
feature observed with {\sl Chandra}: the eastern part has a hard photon index 
($\Gamma\sim$1.2$\pm$0.5), while the south-west part has a softer photon index 
($\Gamma\sim$2.8$\pm$0.7).
\keywords{ISM: supernova remnants: individual: G0.9+0.1 -- X--rays: ISM}
\end{abstract}

\section{Introduction}
  
G0.9+0.1 is the only composite SNR known in the direction of the GC 
(d$\sim$ 10\,kpc, ${\cal{N}}_{\rm H}\sim10^{23}$\,cm$^{-2}$): 
a bright centrally condensed synchrotron nebula (with relatively flat 
spectral index), powered by the loss of rotational energy from the neutron 
star, and a radio shell with a steeper radio spectrum 
(\cite{dporquet-B2:Helfand87}). \cite*{dporquet-B2:Sidoli2000} were able to 
fit the {\sl BeppoSAX} X-ray spectrum of the PWN with a power-law and 
interpret the X-ray emission as non-thermal in origin. The small angular 
extent of the X-ray emission (radius $\sim$ 1\,\arcmin), combined with an 
estimated age of the remnant of a few thousand years, is further evidence 
that the central radio core is powered by a young pulsar ($\sim$ 2,700\,yr; 
\cite{dporquet-B2:Sidoli2000}). Recently, G0.9+0.1 was observed with 
{\sl Chandra} by \cite*{dporquet-B2:Gaensler2001}. 
They found that the synchrotron-emitting pulsar wind nebula (PWN) has 
a clear axial symmetry and a faint X-ray point source lying along the 
symmetry axis possibly corresponds to the pulsar itself. They argued that 
the nebular morphology can be explained in terms of a torus of emission 
in the pulsar's equatorial plane and a jet directed along the pulsar spin 
axis, as is seen in the X-ray nebulae powered by other young pulsars (Crab, 
Vela). 
Here, we present the observations of G0.9+0.1 obtained with {\sl XMM-Newton} 
and we focus on the first spectral analysis of the large and small-scale
structures of the PWN (for more details see \cite{dporquet-B2:Porquet2002}).

\section{Observations and data analysis}

We present in this section the spectral analysis of the PWN using combined 
fits of the MOS and PN data. The observation time after screening rejection 
of the flares are respectively for MOS1 and MOS2 about 15.5\,ksec and 
15.4\,ksec, and 10\,ksec for PN. We use the following camera response 
matrices: m1$\_$medv9q20t5r6$\_$all$\_$15.rsp, 
m2$\_$med\-v9q20t5r6$\_$all$\_$15.rsp, epn$\_$ef20$\_$sdY9$\_$medium.rsp. 
The local background used corresponds to the central MOS\,1 CCD region 
excluding the bright point sources and the central part within a radius 
of 2.6'. We subtract from the source and the local background in our 
pointing, a blank-field observation (kindly provided by David Lumb) at the 
same position in order to take into account the particle background (see 
e.g. \cite{dporquet-B2:Maj2001}).
The normalization between our pointing and the blank-field files is 
determined respectively by the count rate ratio in the 10--12\,keV range 
for the MOS and 12--14\,keV for the PN. The EPIC spectra were binned to 
give S/N\,$\geq$\,3. {\sc xspec v11.1.0} was used for the spectral analysis.
All subsequent errors are quoted at 90$\%$ confidence. 
Abundances are those of \cite*{dporquet-B2:Anders89}. 

\begin{figure}[!th]
\hspace{-2.5cm}
\hspace*{2.5cm}\psfig{figure=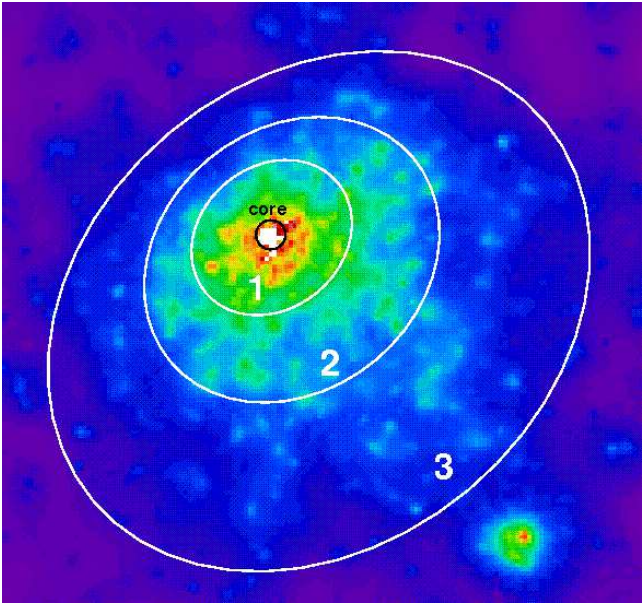,height=8.5cm,angle=-90}
\hspace{-2.5cm}
\caption[]{{\sl XMM-Newton} (EPIC) image in the 3-8\,keV energy band of the PWN 
in G0.9+0.1 obtained with an adaptative smoothing of signal to noise of 5. 
This size of the image is 2.7\,\arcmin\,$\times$\,2.7\arcmin. The different 
large-scale regions used for the spectral analysis are superposed.}
\label{fig:image}
\end{figure}

\section{The overall nebula}

The {\sl XMM-Newton} observation of the supernova remnant G0.9+0.1 shows 
emission from the PWN, but not from the surrounding shell observed in radio. 
The soft X-ray emission expected from the shell is likely not seen due 
to the large interstellar absorption towards the source. 
The EPIC image of the PWN in G0.9+0.1 is presented in Figure~\ref{fig:image}. 
It exhibits a very bright central core, surrounded by extended diffuse 
emission. The extent of this emission is about twice larger than observed 
with {\sl Chandra}, thanks to the high sensitivity of {\sl XMM-Newton}.

For comparison with previous analysis, we fit the spectrum of the overall 
nebula. The region considered corresponds to the sum of the core with 
regions 1, 2 and 3, as is shown in Figure~\ref{fig:image}. The data are 
well fitted by either a power-law or a thermal bremsstrahlung 
(Table~\ref{table:table1}). For the photo-electric absorption, we use the 
cross-sections of \cite*{dporquet-B2:Morrison83}. The parameters 
(${\cal{N}}_{H}$, $\Gamma$) related to the power-law are compatible,
 within the error bars, with the {\sl BeppoSAX} values 
(\cite{dporquet-B2:Sidoli2000}: 
${\cal{N}}_{H}$=10.9$^{+2.4}_{-2.1}$\,cm$^{-2}$, $\Gamma$=1.95$^{+0.33}_{-0.30}$), 
though slightly higher but more tightly constrained. 
In the following, we use the updated cross-sections for X-ray absorption by 
the ISM ({\sc tbabs} in {\sc xspec}) from \cite*{dporquet-B2:Wilms2000}.  
The parameters although in good agreement with the previous ones (see 
Table~\ref{table:table1}), give systematically slightly lower 
${\cal{N}}_{H}$ values using {\sc tbabs} absorption model.
 
\begin{table}[!t]
\begin{center}
\caption[]{
Results of the spectral fits for the overall X-ray PWN (core and regions 1, 2, 
3 in Fig.~\ref{fig:image}). Cross-sections of the interstellar absorption are  
from Morrison \& McCammon (1983; {\sc wabs} in 
{\sc XSPEC}) or from Wilms et al. (2000; {\sc tbabs}).
Uncertainties are quoted at 90$\%$ confidence. The unabsorbed fluxes 
(2--10\,keV) are expressed in $10^{-12}$ erg\,cm$^{-2}$\,s$^{-1}$. }
\label{table:table1}
\begin{tabular}{cccccc}
\hline
\hline
\noalign {\smallskip}
Model          & ${\cal N}_{H}$        & {\small $\Gamma$ or $kT$} & $\chi_{\nu}^{2}$ & F$^{\mathrm \tiny unabs}_{2-10}$ \\
               & {\tiny ($10^{23}$\,cm$^{-2}$)}&      \,\,\,\,\,\,\,\,{\tiny (keV)}&  {\tiny (446 dof)}    &\\
\hline
wabs*PL       & 1.46$^{+0.14}_{-0.13}$      & 2.00$^{+0.19}_{-0.18}$   &   {\small 376.4}   & 5.74   \\
wabs*brems    & 1.33$^{+0.11}_{-0.10}$      & 10.6$^{+4.1}_{-2.2}$    &   {\small 376.1} & 5.11  \\
 wabs*bb      & 0.83$^{+0.09}_{-0.08}$      & 1.73$^{+0.05}_{-0.08}$   &   {\small 389.0}   & 3.63  \\
\hline
tbabs*PL      & 1.39$^{+0.13}_{-0.13}$      & 1.94$^{+0.18}_{-0.18}$   &   {\small 375.2}   & 5.69\\
 tbabs*brems    & 1.27$^{+0.10}_{-0.10}$      & 11.5$^{+4.8}_{-2.6}$    &   {\small 375.0}    & 5.12\\
tbabs*bb      & 0.80$^{+0.09}_{-0.08}$      & 1.75$^{+0.09}_{-0.08}$   &   {\small 387.9}     & 3.65    \\
\hline
\noalign {\smallskip}                       
\hline

\end{tabular}
\end{center}
\end{table}

\section{The large-scale structures}\label{sec:plerion}

\begin{figure}[!t]
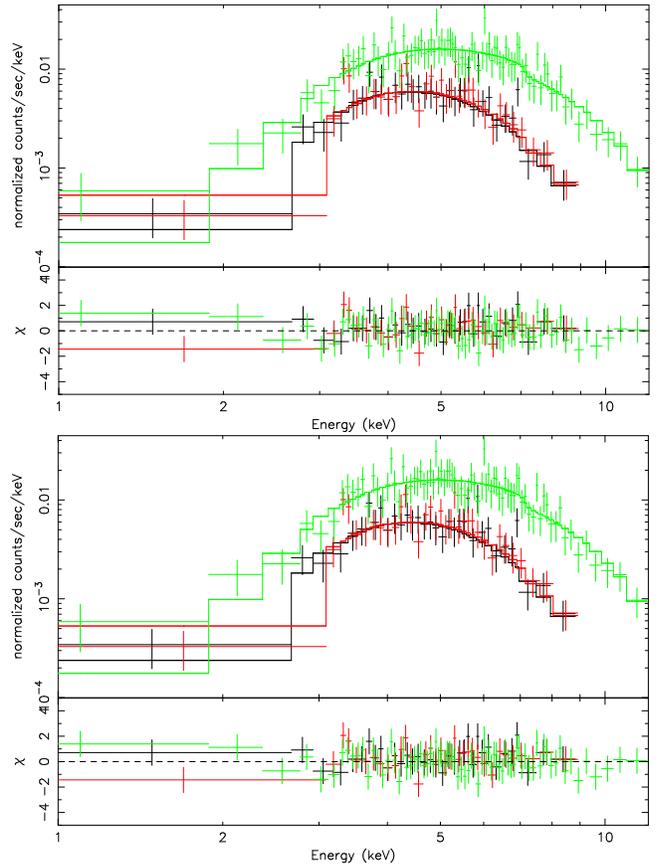

\hspace{-2.5cm}
\hspace{2.4cm}\psfig{figure=dporquet-B2_fig2a.ps,height=8.5cm,angle=-90}\\
\hspace{-1cm}\psfig{figure=dporquet-B2_fig2b.ps,height=8.5cm,angle=-90}
\hspace{-2.5cm}
      \caption[]{Spectral fits of the EPIC data (black for MOS 1, red for MOS2 and green 
for PN) with an absorbed power-law ({\sc tbabs*po}). The inferred parameters are given in 
Table~\ref{table:table2}. {\it Top panel}: Region 1. {\it Bottom panel}: Region 3.}
   \label{fig:spectra}
\end{figure}

The {\sl XMM-Newton} data enables us to study the variation of the spectral 
index within the PWN. In other plerions, a softening of the spectrum has 
been observed toward the outer regions like in 3C 58 
(\cite{dporquet-B2:Torii2000}, \cite{dporquet-B2:Bocchino2001}), G21.5-0.9 
(\cite{dporquet-B2:Slane2000}, \cite{dporquet-B2:Warwick2001}) and IC 443 
(\cite{dporquet-B2:BocchinoBykov2001}). This softening could be explained 
by the shorter lifetime of high energy electrons than for the lower energy 
electrons. \\
To search for this effect, we 
extracted the spectra in four regions, which are displayed in 
Figure~\ref{fig:image}. The observed spectra and best-fit models for the 
regions 1 (top panel) and 3 (bottom panel) are shown, as examples, in 
Figure~\ref{fig:spectra}. The spectrum extends up to almost 12\,keV for the 
PN data, and to about 9\,keV in the MOS data allowing strong constraints 
on the determination of the photon index.
                     
\begin{table}[!b]
\caption{Combined fits of the regions of the PWN in G0.9+0.1. The 
cross-sections of the interstellar absorption are from Wilms et al. (2000). 
Unabsorbed X-ray flux (2-10\,keV) are expressed in
 10$^{-12}$\,ergs\,cm$^{-2}$\,s$^{-1}$. {\it Top}: ${\cal{N}}_{\rm H}$ is 
frozen to the value obtained for the overall PWN (see 
Table~\ref{table:table1}). {\it Bottom}: ${\cal{N}}_{\rm H}$ is a free 
parameter.} 
\begin{tabular}{ccccc}
\hline
\hline
           &  ${\cal{N}}_{\rm H}^{(a)}$    &     $\Gamma$ &  $\chi_{\nu}^{2}$/dof  &   F$_{\mathrm X}^{(b)}$   \\
           & ({\tiny $10^{23}$\,cm$^{-2}$})&               &              &{\tiny (2-10\,keV)}\\
\hline
\hline
core          &1.39                             & 1.12$^{+0.45}_{-0.48}$  & {\small 15.8/18} &  0.25\\
region 1   &1.39                             & 1.46$^{+0.14}_{-0.14}$  & {\small 103.4/146} &  1.79\\
region 2   &1.39                             & 2.02$^{+0.17}_{-0.17}$  & {\small 121.6/136} &  1.65\\
region 3   &1.39                             & 2.42$^{+0.19}_{-0.19}$  & {\small 161.8/173} & 2.02 \\
\hline
\hline
core          & 1.45$^{+0.88}_{-0.66}$     & 1.18$^{+1.17}_{-0.99}$      & {\small 15.8/17}&  0.26\\
region 1   &1.53$^{+0.25}_{-0.22}$  & 1.62$^{+0.30}_{-0.28}$ & {\small 102.3/145} &  1.93\\
region 2   &1.54$^{+0.27}_{-0.23}$  & 2.21$^{+0.38}_{-0.34}$ & {\small 120.6/135}&  1.83\\
region 3   &1.35$^{+0.24}_{-0.21}$  & 2.36$^{+0.39}_{-0.36}$ & {\small 161.7/172} &  1.95\\
\hline
\hline
&\\
\end{tabular}
\label{table:table2}
\end{table}

We fit the spectra of these four regions by absorbed power-laws, fixing 
$\cal{N}_{\rm H}$ to the value obtained for the overall PWN, i.e 
$\cal{N}_{\rm H}$= 1.43\,10$^{23}$\,cm$^{-2}$ 
(see Table~\ref{table:table1}). The best-fit parameters are given in 
Table~\ref{table:table2}. Letting the absorption column densities 
free do not afford significantly better fits, and the values of column 
density $\cal{N}_{\rm H}$ are compatible to within 15$\%$ for the different
regions of the nebula.  

\begin{figure}[!t]
\hspace*{0.3cm}\psfig{figure=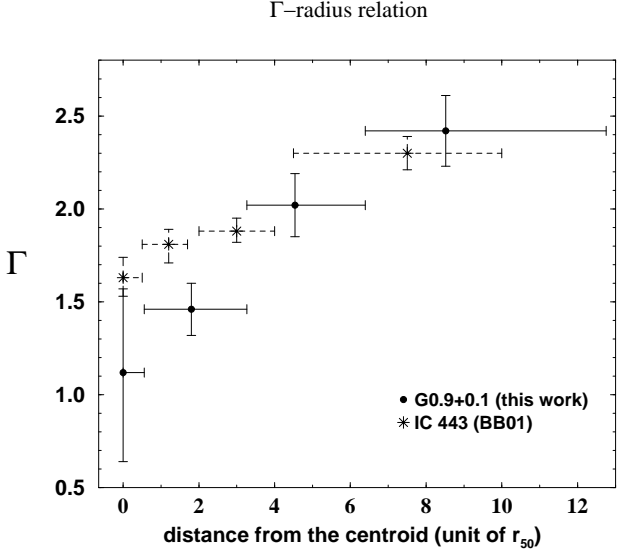,height=8cm,angle=-90}\\
\caption{ Variation of the spectral index versus distance to the 
centroid of the nebula. The X-axis shows the weighted mean distance of the 
pixels in a given region from the centroid expressed in unit of r$_{50}$, the 
radius at which the plerion surface brightness drops by a factor of 2 
(7.15\arcsec~ for G0.9+0.1), as defined in Bocchino \& Bykov (2001). For 
comparison, data for IC\,443 are also plotted (taken from Bocchino \& 
Bykov 2001).}
\label{fig:Gamma-r}
\end{figure}

A clear steepening of the spectrum is observed from the inner part toward 
the outer part of the PWN as is shown in Figure~\ref{fig:Gamma-r}. The 
observed spectral softening with radius in G0.9+0.1 from the core to the 
outskirts of the PWN, is most likely due to the effect of  synchrotron 
radiation losses on the highest energy electrons as they progress through the 
nebula. For comparison we have reported on Figure~\ref{fig:Gamma-r} the 
variation of the spectral index inferred from the observation of IC\,443 
by \cite*{dporquet-B2:BocchinoBykov2001}. The axis unit 
used (r$_{50}$, as defined in Fig.~\ref{fig:Gamma-r} and 
in \cite{dporquet-B2:BocchinoBykov2001}) gives a measure independent 
on the distance to the nebulae. 
Our data appear to show a stronger softening from the core to the outer 
part of the nebula of G0.9+0.1, compared to IC\,443 
(see Fig.~\ref{fig:Gamma-r}), and also to other PWNe such as 3C\,58 and 
G21.5-0.9 (see for comparison Fig.~5 in \cite{dporquet-B2:BocchinoBykov2001}). 
This could be consistent with a stronger magnetic field in the Galactic Center 
region compared to other SNR environments. 

\section{The small-scale structures}

\begin{figure}[!b]
\includegraphics*[height=8.cm]{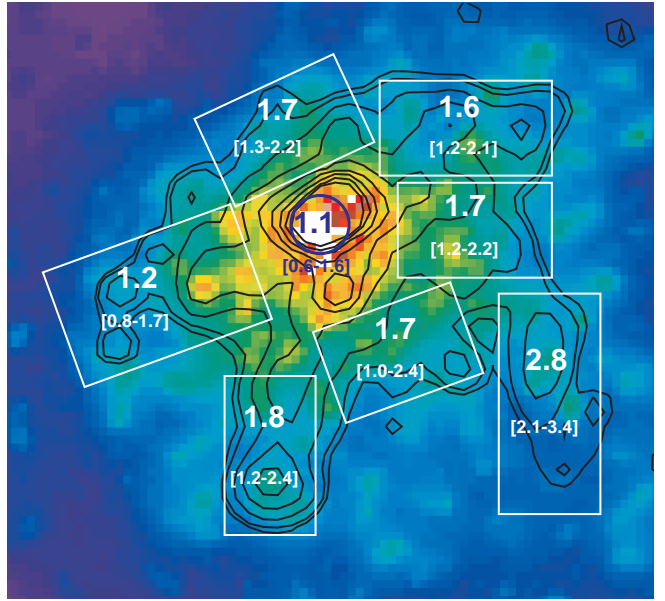}
\caption{
Close-up view with {\sl XMM-Newton} (EPIC) of the center of the nebula in the 
3-8\,keV energy band obtained with an adaptative smoothing of signal to noise of 5. 
The size of the image is 1.7\,\arcmin\,$\times$\,1.3\,\arcmin. The contours 
(in black) corresponds to the {\sl Chandra} observation in the same energy range. 
The white boxes represent different small-scale structures surrounding the 
bright core (dark blue circle), for which the spectral index and 90$\%$ 
confidence uncertainties are given. The position of the structures were 
determined according to the {\sl Chandra} observation (see Gaensler et al. 2001):
 ``East arc'' ($\Gamma \sim$ 1.2),  ``jet'' ($\Gamma \sim$ 1.8), 
and ``South-West arc'' ($\Gamma \sim$ 2.8).
}
\label{fig:gamma}
\end{figure}

The brightest region of the nebula corresponds to a core of angular 
diameter 8\arcsec~ as observed with {\sl XMM-Newton}. The {\sl Chandra} 
observation has resolved it as an elliptical clump of size 
5\arcsec~$\times$~8\arcsec~ (\cite{dporquet-B2:Gaensler2001}). The faint 
point source proposed to be the pulsar by \cite*{dporquet-B2:Gaensler2001} 
is outside this region (about 8\arcsec~ from the center of the clump).
The fact that the hardest spectral index is obtained for the core would 
likely indicate 
that it hosts the pulsar. As proposed by \cite*{dporquet-B2:Gaensler2001}, 
this clump may also be an analog to the knots observed in the Crab nebula 
(\cite{dporquet-B2:Hester95}) and close to the Vela pulsar 
(\cite{dporquet-B2:Pavlov2001}), which exhibit variability in brightness, 
position and morphology. The position and the morphology are however very 
similar in both {\sl XMM-Newton} and {\sl Chandra} observations which are separated from 
only about one month. 
The nature of the bright core is still unclear: a clump, which could be a 
region of separate particle acceleration as in \object{PSR B1509-58} 
(\cite{dporquet-B2:Gaensler2002}), or a very compact region in which the 
 pulsar is embedded?. 

Tracing the spectral variations in the small-scale structures around the core is 
then essential for understanding this particular PWN. Figure~\ref{fig:gamma} 
displays a closer view of the inner region and shows the small-scale 
structures present around the core of PWN G0.9+0.1. For comparison, the 
X-ray contours of the {\sl Chandra} data are superposed. In our data, the 
small-scale structures are not so clearly detected as with {\sl Chandra}  
(jet-like and arc-like features). However using the position of the 
structures seen by {\sl Chandra} and the high sensitivity of {\sl XMM-Newton}, 
we were  for the first time able to carry out spectral analysis of these 
small-scale structure. Determining their spectral index gives an indication of 
the geometry and orientation of the nebula. Figure~\ref{fig:gamma} reports 
the photon spectral index of the corresponding structures obtained with 
combined fits (MOS and PN). The region corresponding to the eastern part 
of the arc-like feature in \cite*{dporquet-B2:Gaensler2001}, shows the 
hardest spectral index ($\Gamma\sim$1.2$\pm$0.5) among all the structures together 
with the core region ($\Gamma\sim$1.1$\pm$ 0.5). On the contrary the region 
corresponding to the south-west part of the arc-like feature, appears to 
have the steepest photon index ($\Gamma\sim$2.8$\pm$0.7)
 over all the PWN. A possible explanation is that the east arc is pointing 
towards the observer and that its spectral hardness is due to the relativistic 
beaming or Doppler boosting of the electrons, while the south-west arc is 
the opposite part of the torus. The region associated to the jet-like feature 
in \cite*{dporquet-B2:Gaensler2001} does not exhibit a harder spectrum as 
suggested by their hardness ratio.

\section{Conclusion}

\indent We report {\sl XMM-Newton} observations of the X-ray Pulsar Wind Nebula (PWN) 
inside the composite supernova remnant G0.9+0.1 (\cite{dporquet-B2:Helfand87}), 
which was convincingly detected for the first time in X-rays by {\sl BeppoSAX} (\cite{dporquet-B2:Mereghetti98}).\\
\indent The large-scale image of the PWN G0.9+0.1 shows a very bright 
central core, surrounded by relatively large-scale diffuse emission with 
a much spatial extent than is apparent for the recent {\sl Chandra} images. 
Thanks to its high sensitivity, {\sl XMM-Newton} allows for the first time a spectral analysis 
of the large and small-scale structures 
(see also \cite{dporquet-B2:Porquet2002}). Our spectral analysis combining the 
MOS and PN data offers a clear indication of a softening of the spectrum 
 with distance from the centroid (related to the finite lifetime of 
the synchrotron emitting electrons), as observed in other known X-ray 
plerions. 
Even if in our data, the small-scale structures are not so clearly detected 
as in the {\sl Chandra} data (jet-like and arc-like features),
 we were able for the first time to carry out their spectral analysis.
Two of the structures show distinct spectral indexes from the general trend 
(from $\Gamma\sim$1.6$\pm$0.5 to $\Gamma\sim$1.8$\pm$0.6) : 
the eastern part of the arc-like feature has a 
very hard photon index ($\Gamma\sim$1.2$\pm$0.5), while the south-west part of the 
arc-like feature has a softer photon index ($\Gamma\sim$2.8$\pm$0.7).\\
\indent The study of G0.9+0.1 provides further clues concerning the processes by which pulsars
connect with their environment and also illustrates the impact 
of a strong ambient magnetic field as in the Galactic Center region.

\begin{acknowledgements}

D.P. acknowledges grant support from the ``Institut National des Sciences de l'Univers'' and from the
``Centre National d'Etudes Spatial'' (France). 
\end{acknowledgements}

\end{document}